\documentclass[11pt]{report}
 
\usepackage{setspace}
\usepackage[margin=1.5truein]{geometry} %1 inch margins
\usepackage[utf8]{inputenc}
\usepackage{authblk} % Does the authorship with superscript thingie
\usepackage{hyperref} % adds urls
\usepackage{color,soul} % Makes realistic looking highlighting
\usepackage{textcomp} % adds things like degrees and mu and stuff
\usepackage{graphicx} %for including graphics
\usepackage{booktabs} %for nice looking latex tables
\usepackage{array}
\usepackage{multirow}
\usepackage{amssymb} %for math
\usepackage{mathtools}
\usepackage{textcomp}
\usepackage{amsbsy}

\usepackage{xr} %External referencing.. reference across files
\usepackage[T1]{fontenc}
\raggedright %and ragged right margin

\usepackage[textsize=tiny, backgroundcolor=yellow]{todonotes}

\newcommand{\reals}[0]{\rm I\!R}
\newcommand{\bfx}[0]{\mathbf{x}}
\newcommand{\bfW}[0]{\mathbf{W}}
\newcommand{\bfr}[0]{\mathbf{r}}
\newcommand{\bfv}[0]{\mathbf{v}}
\newcommand{\bfX}[0]{\mathbf{X}}
\newcommand{\bfR}[0]{\mathbf{R}}
\newcommand{\bfxi}[0]{\boldsymbol{\xi}}

\DeclarePairedDelimiter\norm{\lVert}{\rVert}%

\title{Automatically tracking neurons in a moving and deforming brain}

\author[1,2]{Jeffrey P.~Nguyen} 
\author[3]{Ashley N.~Linder} 
\author[2, \#]{George S.~Plummer} 
\author[1,2]{Joshua W.~Shaevitz}
\author[1,3,*]{Andrew M.~Leifer}

\affil[1]{Department of Physics, Princeton University, Princeton, New Jersey, United States of America}
\affil[2]{Lewis-Sigler Institute for Integrative Genomics, Princeton University, Princeton, New Jersey, United States of America}
\affil[ \#]{Current Address: Tufts University School of Medicine, Boston, Massachusetts, United States of America}
\affil[3]{Princeton Neuroscience Institute, Princeton University, Princeton, New Jersey, United States of America}
\affil[*]{Corresponding author, E-mail: \href{mailto:leifer@princeton.edu}{leifer@princeton.edu} (AML)}

\begin{document}
\maketitle

\begin{abstract}
Advances in optical neuroimaging techniques now allow neural activity to be recorded with cellular resolution in awake and behaving animals.  Brain motion in these recordings pose a unique challenge. The location of individual neurons must  be tracked in 3D over time  to accurately  extract single neuron activity traces. Recordings from small  invertebrates like \textit{C. elegans} are especially challenging because they  undergo very large brain motion and deformation during animal movement. Here we present an automated computer vision pipeline  to reliably track populations of neurons with single neuron resolution in  the  brain of a freely moving \textit{C. elegans} undergoing large motion and deformation. 3D volumetric fluorescent images of the animal's brain are straightened, aligned and registered, and the locations of neurons in the images are found via segmentation. Each neuron is then assigned an identity using a new time-independent machine-learning approach we call Neuron Registration Vector Encoding. In this approach, non-rigid point-set registration is used to match each segmented neuron in each volume with a set of reference volumes taken from throughout the recording. The way each neuron  matches with the references defines a feature vector which is clustered to assign an identity to each neuron in each volume.  Finally, thin-plate spline interpolation is used to correct errors in segmentation and check consistency of assigned identities. The Neuron Registration Vector Encoding approach proposed here  is uniquely well suited for tracking neurons in brains undergoing large deformations. When applied to whole-brain calcium imaging recordings in freely moving \textit{C. elegans}, this analysis  pipeline located 150 neurons for the duration of an 8 minute recording and consistently found more neurons more quickly than manual or semi-automated approaches.
\end{abstract}

\section{Author Summary}

Computer algorithms for identifying and tracking neurons in images of a brain have struggled to keep pace with rapid advances in neuroimaging. In small transparent organism like the nematode \textit{C. elgeans}, it is now possible to record neural activity from all of the neurons in the animal's head with single-cell resolution as it crawls. A critical challenge is to identify and track each individual neuron as the brain moves and bends.  Previous methods required large amounts of manual human annotation. In this work, we present a fully automated algorithm for neuron segmentation and tracking in freely behaving \textit{C. elegans}. Our approach uses non-rigid point-set registration to construct feature vectors describing the location of each neuron relative to other neurons and other volumes in the recording. Then we cluster feature vectors in a time-independent fashion to track neurons through time. This new approach works very well when compared to a human.

\section{Introduction}
Optical neural imaging has ushered in a new frontier in  neuroscience that seeks to understand how neural activity generates animal behavior by  recording from large populations of neurons at cellular resolution in awake and behaving animals.  Population recordings  have now been used to elucidate mechanisms behind zebra finch song production \cite{picardo2016}, spatial encoding in mice \cite{Rickgauer2014}, and limb movement in primates \cite{maynard1999}. When applied to small transparent organisms, like  \textit{Caenorhabditis elegans} \cite{Kato2015}, \textit{Drosophila} \cite{Li2016}, and zebrafish \cite{Prevedel2014}, nearly every neuron in the brain can be recorded, permitting the study of whole brain neural dynamics at cellular resolution. 
 
Methods for segmenting and tracking neurons have struggled to keep up as new imaging technologies now record from more neurons over longer times in environments with greater motion. Accounting for brain motion in particular has become a major challenge, especially in recordings of unrestrained animals. Brains in motion undergo translations and deformations in 3D that make robust tracking of individual neurons very difficult. The problem is compounded in invertebrates like \textit{C. elegans} where the head of the animal is  flexible and deforms greatly. If left unaccounted for, brain motion not only prevents tracking of neurons, but it can also  introduce  artifacts that mask the true neural signal. In this work we propose an automated approach  to segment and track neurons in the presence of dramatic brain motion and deformation. Our approach is optimized for calcium imaging in unrestrained \textit{C. elegans}.  

% Measuring a larger proportion of the neurons in an animal allows for the combination the neural activity with neural connectivity measurements in order to understand how the brain encodes information and behavior\cite{Bargmann2013}. 

Neural activity can be imaged optically with the use of genetically encoded calcium sensitive fluorescent indicators, such as GCaMP6s used in this work \cite{Chen2013}. Historically calcium imaging was often   conducted in head-fixed or anesthetized animals  to avoid challenges involved with imaging moving samples \cite{Kato2015, harvey2009intracellular,Ahrens2012}. Recently, however, whole-brain imaging was demonstrated in freely behaving \textit{C. elegans} \cite{Nguyen2016, Venkatachalam2016}. \textit{C. elegans} are a small transparent nematode, approximately 1mm in length, with a compact nervous system of only 302 neurons. About half of the neurons  are located in the animal's head, which we refer to as its brain. 

%This worm's optical transparency and compact nervous system make it possible  to image the entire brain of the animal non-invasively while the animal moves unrestrained. 

Analyzing fluorescent images of moving and deforming brains requires algorithms to detect neurons across time and extract fluorescent signals in 3D. Several strategies exist for tracking neurons in volumetric recordings. One approach is to find correspondences between neuron positions in consecutive time points, for example, by applying a distance minimization, and then stitching these correspondences together through time \cite{Crocker1996}. This type of time-dependent tracking requires that neuron displacements for each time step are less than the distance between neighboring neurons, and that the neurons remain identifiable at all times. If these requirements break down, even for only a few time points, errors can quickly accumulate.   Other common methods, like independent component analysis (ICA) \cite{Mukamel2009} are also exquisitely sensitive to motion and as a result they have not been successfully applied to recordings with large brain deformations.   

Large inter-volume motion arises when the recorded image volume acquisition rate is too low compared to animal motion.  Unfortunately, large inter-volume brain motion is likely to be a prominent feature of whole-brain recordings of moving brains for the foreseeable future. In all modern imaging approaches there is a fundamental tradeoff between the following attributes: acquisition rate (temporal resolution), spatial resolution, signal to noise, and the spatial extent of the recording. As recordings seek to capture  larger brain regions at single cell resolution, they necessarily compromise on temporal resolution. For example, whole brain imaging in freely moving \textit{C. elegans} has  only been demonstrated at slow acquisition rates because of the requirements to scan the entire brain volume and expose each slice for sufficiently long time. At these rates, a significant amount of motion is present between image planes within a single brain volume. Similarly, large brain motions also remain between sequential volumes.  Neurons can move the entire width of the worm's head between sequential volumes when recording at 6 brain-volumes per second, as in \cite{Nguyen2016}. In addition to motion, the brain also bends and deform as it moves. Such changes to the brain's conformation greatly alter the pattern of neuron positions making constellations of neurons difficult to compare across time. 

To account for this motion, previous work that measured neural activity in freely moving \textit{C. elegans} required either large amounts of manual annotation as reference data for comparison \cite{Venkatachalam2016} or required a human user to supervise and correct semi-automated  algorithms for each and every neuron-time point \cite{Nguyen2016}.  This level of manual annotation becomes impractical as the length of recordings and the number of neurons increases. For example, 10 minutes of recorded neural activity from \cite{Nguyen2016},  had over 360,000 neuron time points and required over 200 person-hours of manual annotation. Here, we introduce a new  time-independent algorithm that uses machine learning to automatically segment and track all neurons in the head of a freely moving animal. We call this technique Neuron Registration Vector Encoding, and we use it to extract neural signals in unrestrained \textit{C. elegans} expressing the calcium indicator GCaMP6s and the fluorescent label RFP.

\begin{figure}
	\begin{center}	
		 \includegraphics[width=0.5\textwidth]{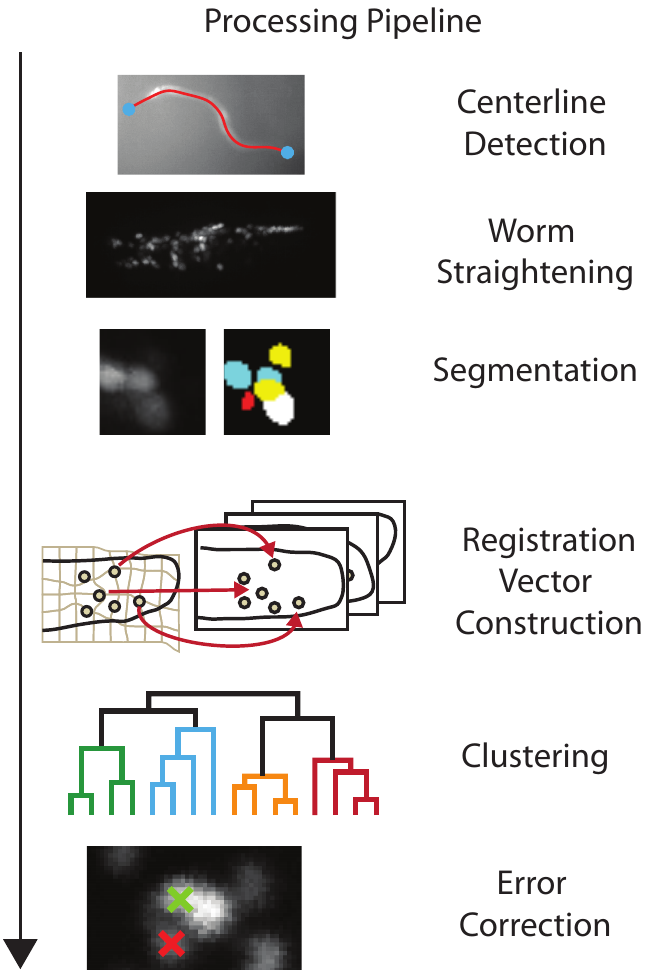} 
	\caption{\label{flow} Schematic of analysis pipeline to segment and track neurons through time and extract their neural activity in a deforming brain. Neurons are labeled with calcium insensitive red fluorescent proteins, RFP, and calcium sensitive green fluorescent proteins, GCaMP.  Videos of the animal's behavior and volumetric fluorescent images of the animal's brain serve as input to the pipeline.  The algorithm detects all neurons in the head and produces tracks of the neural activity across time as the animal moves.} 
	\end{center}
\end{figure}

\section{Results}

\subsection{Overview of neuron tracking analysis}
We introduce a method to track over 100 neurons in the brain of a freely moving \textit{C. elegans}.   The analysis pipeline is made of five modules and an  overview  is shown in Figure \ref{flow}. The first three modules, ``Centerline Detection,''  ``Straightening'' and ``Segmentation,'' collectively assemble the individually recorded planes into a sequence of 3D volumes  and  identify  each neuron's location in each volume.  The next two modules,  ``Registration Vector Construction'' and ``Clustering,'' form the core of the method and represent a significant advance over previous approaches. Collectively, these two modules are called ``Neuron Registration Vector Encoding".  The ``Registration Vector Construction''  module leverages information from across the entire recording in a time-independent way to generate feature vectors that characterize every neuron at every time point in relation to a repertoire of brain confirmations. The ``Clustering'' module then clusters these feature vectors to assign a consistent identity to each neuron across the entire recording. A final module corrects for errors the original segmentation. The implementation and results of this approach are described below.

\subsection{Recording of whole-brain calcium activity and body posture in moving animal}

Worms expressing the calcium indicator GCaMP6s and a calcium-insensitive fluorescent protein RFP in the nuclei of all neurons were imaged during unrestrained behavior in a custom 3D tracking microscope, as described in \cite{Nguyen2016}.  Two recordings are presented in  this work:  a new 8 minute recording of an animal of of strain AML32  and a previously reported 4 minute recording of strain first described in  \cite{Nguyen2016}. 

\begin{figure}
	\begin{center}	
		 \includegraphics[width=\textwidth]{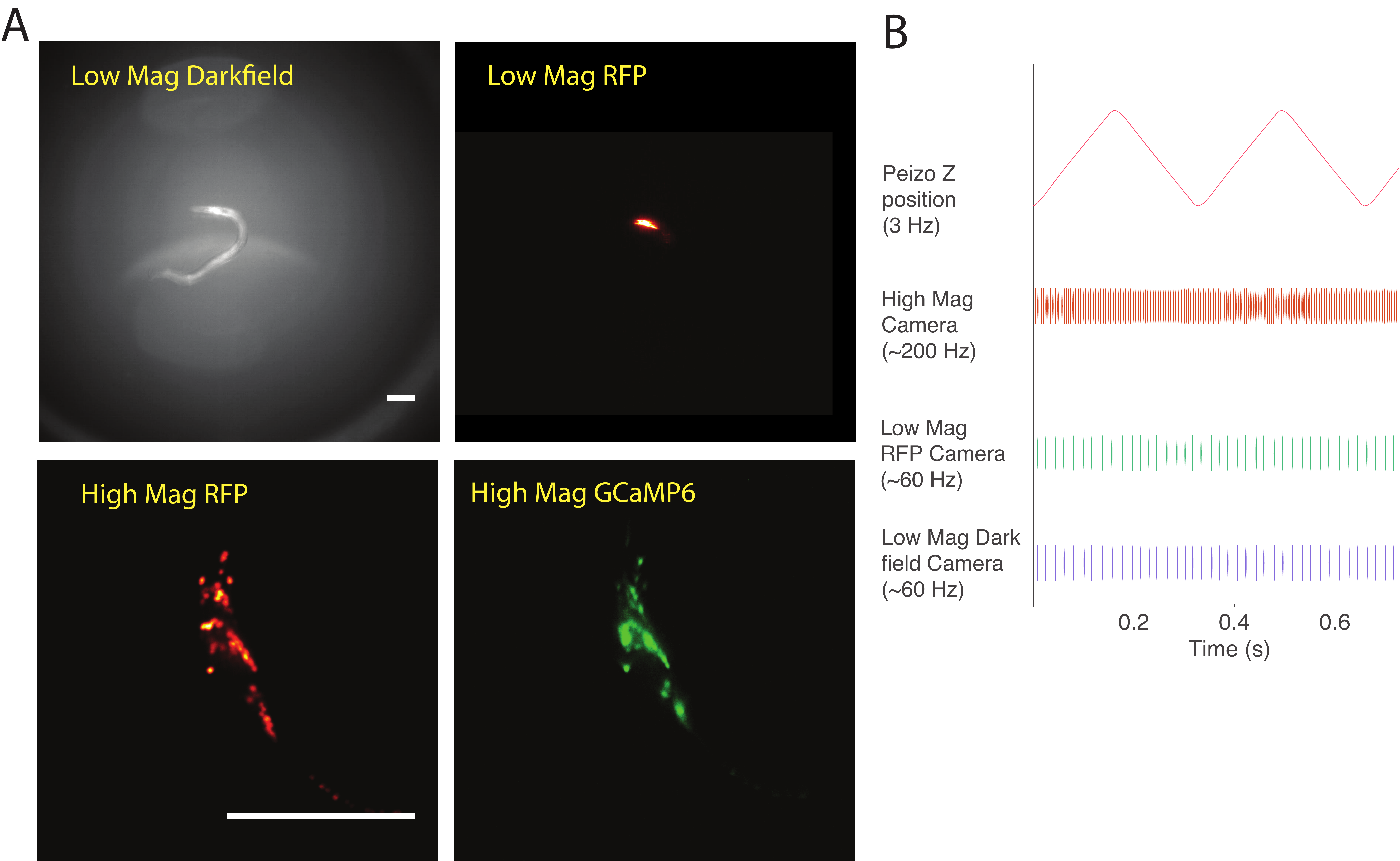}
	\caption{\label{Optics}(A) Example images from all four video feeds from our imaging system. Both scale bars are 100$\mu$m (B)A schematic illustrating  the timings from all the devices that run in open loop in our imaging setup. The camera that collects high magnification images captures at 200Hz. The two low magnification images capture at 60Hz, and the focal plane moves up and down in a 3 Hz triangle wave. The cameras are synchronized post-hoc using a camera flash and each image is assigned a timestamp on a common timeline for the purposes of analysis.}
	\end{center}
\end{figure}

The signal of interest in both recordings is the green fluorescence intensity from GCaMP6s in each neuron. Red fluorescence from the RFP protein serves as a reference for locating and tracking the neurons.  The microscope provides four raw image streams that serve as inputs for our neural tracking pipeline, seen in Fig \ref{Optics}A.   They are: (1) low-magnification dark-field images of the animal's body posture (2) low-magnification fluorescent images of the animal's brain (3) high-magnification green fluorescent images of single optical slices of the brain showing GCaMP6s activity and (4) high-magnification red fluorescent images of single optical slices of the brain showing the location of RFP.  The animal's brain is kept centered in the field of view by realtime feedback loops that adjust a motorized stage to compensate for the animal's crawling. To acquire volumetric information, the high magnification imaging plane scans back and forth along the axial dimension, $z$,  at 3 Hz as shown in Fig \ref{Optics}B, acquiring roughly 33 optical slices per volume, sequentially, for 6 brain-volumes per second. The animal's continuous motion causes  each volume to be arbitrarily sheared in $z$.  Although the image streams  operate at different volume acquisition rates and on different clocks, they are later synchronized by simultaneous light flashes and given a timestamp on a common timeline. Each of the four imaging streams are aligned to each other in software using affine transformations found by imaging fluorescent beads.

\subsection{Centerline detection and gross brain alignment}
The animal's posture contains information about the brain's orientation and about any deformations arising from the animal's side-to-side head swings. The first step of the pipeline is to extract the centerline that describes the animal's posture. Centerline detection in \textit{C. elegans} is an active field of research. Most algorithms use intensity thresholds to detect the worm's body and then use binary image operations to extract a centerline \cite{Stephens2008, Peng2008, stephens2016}. Here we use an open active contour approach \cite{deng2013, kass1988} to extract the centerline from dark field images with  modifications to account for cases when the worm's body crosses over itself as occurs during so-called ``Omega Turns.''  In principle any method, automated or otherwise, that detects the centerlines should be sufficient.  At rare times where the worm is coiled and the head position and orientation cannot be determined automatically, the head and the tail of the worm are manually identified.

\begin{figure}
	\begin{center}	
		 \includegraphics[width=\textwidth]{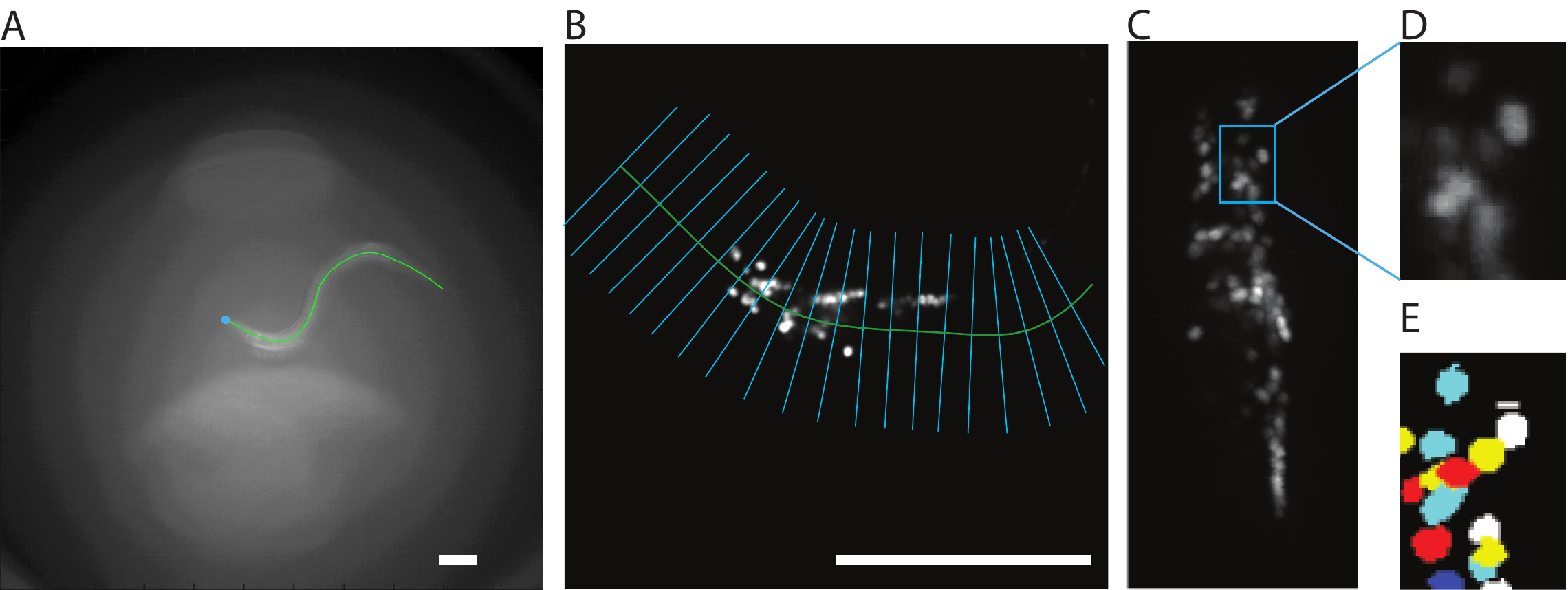}
	\caption{\label{StraighteningFigure} (A) Centerlines are detected from the low magnification dark field images. The centerline is shown in green and the tip  of the worm's head is indicated by a blue dot. (B) The centerline found from the low magnification image is overlaid on the high magnification RFP images. The lines normal to the centerline, shown in blue, are used to straighten the image. All scale bars are 100 $\mu$m.(C) A maximum intensity projection of the straightened volume is shown. Individual neuronal nuclei are shown (D) before  and (E) after segmentation.}
	\end{center}
\end{figure}

The animal's centerline allows us to correct for gross changes in the worm's position, orientation, and conformation (Fig \ref{StraighteningFigure}a). We use the centerlines determined by the low magnification behavior images to straighten the high magnification images of the worm's brain. An affine transform must be applied to the centerline coordinates to transform them from the dark field coordinate system into the coordinate system of the high magnification images. Each image slice of the worm brain is straightened independently to account for motion within a single volume. The behavior images are taken at a lower acquisition rate than the high magnification brain images, so a linear interpolation is to used obtain a centerline for each slice of the brain volume. In each slice, we find the tangent and normal vectors at every point of the centerline (Fig \ref{StraighteningFigure}b). The points are interpolated with a single pixel spacing along the centerline to preserve the resolution of the image. The image intensities along each of the normal directions are interpolated and the slices are stacked to produce a straightened image in each slice (Fig \ref{StraighteningFigure}c). In the new coordinate system, the orientation of the animal is fixed and the worm's bending is greatly suppressed. We further reduce shearing between slices using standard video stabilization techniques \cite{tordoff2002guided}. Specifically, bright-intensity peaks in the image are tracked between neighboring image slices and affine transformations are calculated between each slice of the volume. All slices are registered to the middle slice by applying these transformations sequentially throughout the volume. Each slice would undergo  transformations for every slice in between it and the middle slice to correct shear throughout the volume. A final rigid translation is required to align each volume to the initial volume. The translations are found by finding an offset that maximizes the cross-correlation between each volume and the initial volume. 

\subsection{Segmentation}

Before neuron identities can be matched across time, we must first segment the individual neurons within a volume to recovers each neuron's size, location, and brightness (Fig \ref{StraighteningFigure}d,e). Many algorithms have been developed to segment neurons in a dense region \cite{Lin2003,Toyoshima2016}. We segment the neurons by finding volumes of curvature in fluorescence intensity. We compute the 3D Hessian matrix at each point in space and threshold for points where all of the three eigenvalues of the Hessian matrix are negative. In order to further divide regions into objects that are more likely to represent neurons, we use a watershed separation on the distance transform of the thresholded image. The distance transform is found by replacing each thresholded pixel with the Euclidean distance between it and the closest zero pixel in the thresholded image. Image blurring from animal motion poses a challenge for segmentation. We allow for some noise and error in the segmentation because we will have the opportunity to automatically correct many of these errors later in the pipeline. 

\subsection{Neuron registration vector construction}

\begin{figure}
	\begin{center}	
		 \includegraphics[width=\textwidth]{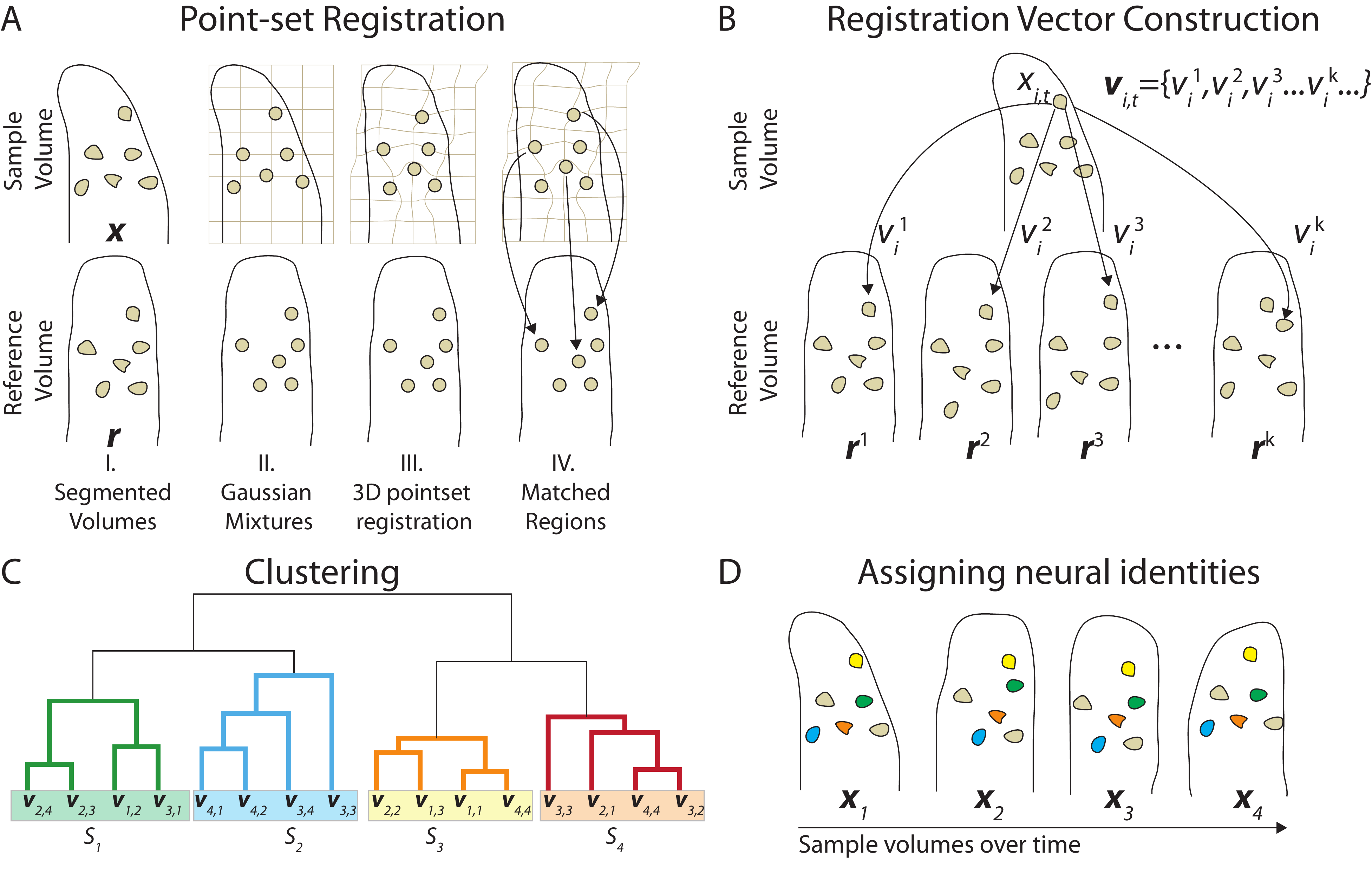}
	\caption{\label{Registration} Schematic illustration of neuron registration vector encoding. (A) The registration between a sample volume and a single reference volume is done in several steps. I. The image is segmented into regions corresponding to each of the neurons.  II. The image is represented as a Gaussian mixture, with a single Gaussian for each segmented region. The amplitude and the standard deviation of the Gaussians are derived from the brightness and the size of the segmented regions. III. Non-rigid point-set registration is then used to deform the sample points to best overlap the reference point-set. IV. Neurons from the sample and the reference point-sets are paired by minimizing distances between neurons. (B) Neuron registration vectors are constructed by assigning a feature vector $\bfv_{i,t}$ to each neuron $x_{i,t}$ in a sample volume $\bfx_t$  by performing the registration between the sample volume and a set of 300 reference volumes, each denoted by $\bfr^k$. Each registration of the neuron results in a neuron match, $v_i^k$, and the set of matches becomes the feature vector $\bfv_{i,t}$. (C) The vectors from all neuron-times, $\bfv_{i,t}$, are hierarchically clustered. The same neuron found at different times will have a similar set of features and therefore will contain the same neuron found at different times. Real matches occur in a high dimensional space. Only two dimensions are illustrated here for clarity. Each of the feature vector is assigned a cluster, and the cluster labels are given by $S$. (D) The clustering of the feature vectors shown in (C) assigns an identity to each of the neurons in every volume. This allows us to track the neurons across different volumes of the recording.}
	\end{center}
\end{figure}

Extracting neural signals requires the ability to match neurons found at different time points. Even after gross alignment and straightening, neurons in our images are still subject to local nonlinear deformations and there is significant movement of neurons between volumes. Rather than tracking through time, the neurons in each volume are characterized based on how they match to neurons in a set of  reference volumes. Our algorithm compares constellations of neurons in one volume to unannotated reference volumes and assigns correspondences or ``matches'' between the neurons in the sample and each reference volume. We modified a point set registration algorithm developed by Jian and Vemuri \cite{Jian2005} to do this (Fig \ref{Registration}a). The registration algorithm represents two point sets, a sample point-set denoted by $\bfX = \{\bfx_i\}$ and a reference point-set indicated by  $\bfR=\{\bfr_i\}$,  as Gaussian mixtures and then attempts to register them by deforming space to minimize the distance between the two mixtures. Here, each neuron is modeled by a 3D Gaussian with uniform covariance. Since we are matching images of neurons rather than just points, we can use the additional information from the size and brightness of each neuron. We add this information to the representation of each neuron by adjusting the amplitude and standard deviation of the Gaussians. The Gaussian mixture representation of an image is given by,
\begin{equation}
f(\bfxi, \bfX)=\sum_i A_i\exp \bigg(- \frac{\norm{\bfxi-\bfx_i}^2}{2(\lambda \sigma_i)^2}\bigg), 
\end{equation}
where $A_i$, $\bfx_i$, and $\sigma_i$ are the amplitude, mean, and standard deviation of the $i$-th Gaussian. These parameters are derived from the brightness, centroid, and size of the segmented neuron, while $\bfxi$ is the 3D spatial coordinate. A scale factor $\lambda$ is added to the standard deviation to scale the size of each Gaussian. This will be used later during gradient descent. The sample constellation of neurons is then represented by the Gaussian mixture $f(\bfxi,\bfX)$. Similarly, the reference constellation's own neurons is represented as a $f(\bfxi, \bfR)$.

To match a sample constellation of neurons  $\bfX$ with a reference constellation of neurons  $\bfR$, we use the non rigid transformation $u: \reals^3\mapsto \reals^3 $. The transformation $u$ maps $\bfX$ to $u[\bfX]$ such that the $L_2$ distance between $f(\bfxi,u[\bfX])$ and $f(\bfxi,\bfR)$ is minimized with some constraint on the amount of deformation. This can be written as an energy minimization problem, with the energy of the transformation, $E(u)$, written as   
\begin{equation}
E(u)=\int \big[f(\bfxi,u[\bfX]) -f(\bfxi,\bfR)\big]^2 d\bfxi +  E_{\mathrm{Deformation}}(u).
\label{Energy}
\end{equation}

Note that the point-sets $\bfX$ and $\bfR$ are allowed to have different numbers of points. We model the deformations as a thin-plate spline (TPS). The transformation equations and resulting form of $E_{\mathrm{deformation}}(u)$ is shown in the methods. The minimization of $E$ is found by gradient descent. Since the energy landscape has many local minima, we initially chose a large scale factor, $\lambda$, to increase the size of each Gaussian and smooth over smaller features. During gradient descent, $\lambda$ is decreased to better represent the original image. After the transformation, sample points are matched to reference points by minimizing distances between assigned pairs \cite{Crocker1996}. The matching is not greedy, and neurons in the sample that are far from any neurons in the reference are not matched. A neuron at $\bfx_i$ is assigned a match  $v_i$ to indicate which neuron in the set $\bfR$ it was matched to. 
For example if $\bfx_i$ matched with $\bfr_j$ when $\bfX$ is registered to $\bfR$, then $v_i = j$. If $\bfx_i$ has no match in $\bfR$, then $v_i=\emptyset$.

The modified non-rigid point set registration algorithm described above allows us to compare one constellation of neurons to another. In principle, neuron tracking could be achieved by registering the constellation of neurons at each time-volume to a single common reference. That approach is susceptible to failures in non-rigid point set registration. Non-rigid point-set registration works well when the conformation of the animal in the sample and the reference are similar, but it is unreliable when there are large deformations between the sample and the reference, as happens with some regularity in our recordings. In addition, this approach is especially sensitive to any errors in segmentation, especially in the reference. An alternative approach would  be to sequentially register neurons in each time volume to the next time-volume. This approach, however, accumulates even small errors and quickly becomes unreliable.  Instead of either of those approaches, we use registration to compare the constellation of neurons at each time volume to a set of reference time-volumes that span a representative space of brain conformations (Fig \ref{Registration}b), as described below. 

 The constellation of neurons at a particular time in our recording is given by $\bfX_t$, and the position of the $i$-th neuron at time $t$ is denoted by $\bfx_{i,t}$. We select a set of $K$ reference constellations, each from a different time volume $\bfX_t$ in our recording, so as to achieve a representative sampling of the many different possible brain conformations the animal can attain. These $K$ reference volumes are denoted by $\{\bfR^1, \bfR^2, \bfR^3, ..., \bfR^K\}$. For simplicity, we use 300  volumes spaced evenly through time as our reference constellations. Each $\bfX_t$ is separately matched with each of the references, and each neuron in the sample, $\bfx_{i,t}$, gets a set of matches $\bfv_{i,t}=\{v_{i,t}^1, v_{i,t}^2, v_{i,t}^3,..v_{i,t}^K\}$, one match for each of the $K$ references. This set of matches is a feature vector which we call a Neuron Registration Vector. It describes the neuron's location in relation to its neighbors when compared with the set of references. This vector can be used to identify neurons across different times.
 
\subsection{Clustering registration vectors}
The neuron registration vector  provides information about that neuron's position relative to its neighbors, and how that relative position compares with many other reference volumes. A neuron with a particular identity will match similarly to the set of reference volumes and thus that neuron will have similar neuron registration vectors over time. Clustering similar registration vectors allows for the identification of that particular neuron across time (Fig \ref{Registration}c,d).

%%%%%%%%%%%%%%
To illustrate the motivation for clustering, consider a neuron with identity $s$ that is found at different times in two sample constellations $\boldsymbol{X}_1$ and  $\boldsymbol{X}_2$. When $\boldsymbol{X}_1$ and  $\boldsymbol{X}_2$ have similar deformations, the neuron $s$ from both constellations will be assigned the same set of matches when registered to the set of reference constellations, and as a result the corresponding neuron registration vectors $\bfv_1$ and $\bfv_2$ will be identical. This is true even if the registration algorithm itself fails to correctly match neuron $s$  in the sample to its true neuron $s$ in the reference.  As the deformations separating $\boldsymbol{X}_1$ and $\boldsymbol{X}_2$ become larger, the distance between the feature vectors $\bfv_1$ and $\bfv_2$  also becomes larger. This is because the two samples will be matched to different neurons in some of the reference volumes as each sample is more likely to register poorly with references that are far from it in the space of deformations.  

Crucially, the reference volumes consist of instances of the animal in many different deformation states. So while errors in registering some samples will exist for certain reference, they do  not  persist across all references, and thus do not effect the entire feature vector.  For the biologically relevant deformations that we observe, the distance between $\bfv_1$ and $\bfv_2$  will be smaller if both are derived from neuron $s$ than compared to the distance between $\bfv_1$ and $\bfv_2$ if they were derived from $s$ and another neuron.  We can therefore cluster the feature vectors to produce groups that consist of the same neuron found at many different time points. 

%%%%%%%%%%%%%%%

%The neurons in $\boldsymbol{A}$ can be correctly matched with the corresponding neurons in $\boldsymbol{B}$ by finding pairs of neurons with identical fingerprints. This pairing is robust even to errors in registration because registration affect the registration of neurons in both $\boldsymbol{A} \rightarrow \boldsymbol{C}$ and $\boldsymbol{B} \rightarrow \boldsymbol{C}$ in the same way. 

%When two different sample constellations are used, $\boldsymbol{A}\neq\boldsymbol{B}$, some of the fingerprints achieved from registering  $\boldsymbol{A} \rightarrow \boldsymbol{B}$ and $\boldsymbol{A} \rightarrow \boldsymbol{C}$ may no longer be identical due to \hl{errors in  the registration}\todo{should the error be in the registration itself? the sample or the reference?}. When the deformations between two constellations is large,  the energy minimization can end in one of the many local minima in the landscape. and  when the deformations are between the two constellations is large, the right solution may not be achieved. 

The list of neuron registration vectors from all neuron at all times, $\{\bfv_{i,t}\}$, is hierarchically clustered. Each match in the vectors, $v_{i,t}^k$, is represented as a binary vector of 0s with a 1 at the $\bfv_{i}^{k\mathrm{-th}}$ position.  The size of the vector is equal to the number of neurons in $\bfR^k$. The feature vector $\{\bfv_{i,t}\}$ is the concatenation of all of the binary vectors from all matches to the $K$ reference constellations. The correlation distance, $1 - \mathrm{corr}(\bfv_m,\bfv_n)$, was used as the pair wise distance metric for clustering. Clusters were checked to ensure that at most one neuron was represented from each time. Clusters containing neurons from less than 40\% of the volumes were removed. Each cluster is assigned a label $\{S_1, S_2, S_3,...\}$ which uniquely identifies a single neuron over time, and each neuron at each time $\bfx_{i,t}$ is given an identifier $s_{i,t}$ which corresponds to which cluster that neuron-time belongs to. Neurons that are not assigned to one of these clusters are removed because they are likely artifactual or represent a neuron that is segmented too poorly for inclusion. 

 %\hl{The number of clusters is not set, so that the algorithm determines how many neurons are present in our recordings.} 

\subsection{Correcting errors in tracking and segmentation }

\begin{figure}
	\begin{center}	
		 \includegraphics[width=\textwidth]{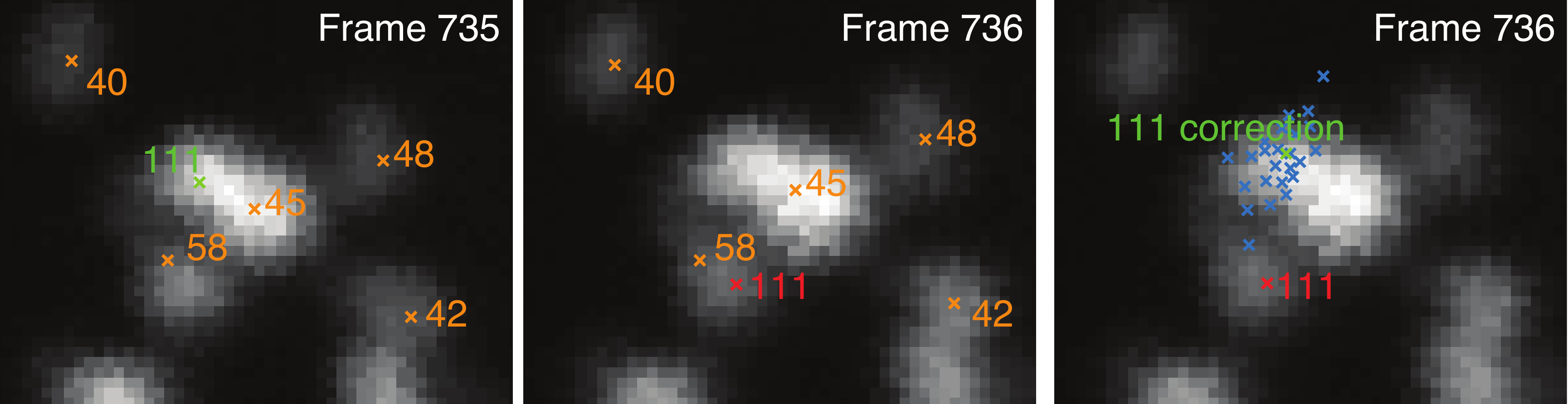}
	\caption{\label{correction} Example of consensus voting to correct a misidentified neuron. In volume 735, neuron \#111 is found successfully and is indicated in green. In volume 736, however, the neuron is misidentified, shown in red. During the correction phase, all other time points vote for what the position of neuron \#111 should be assuming a thin-plate spline deformation, indicated by the cloud of blue points. Since the initial estimate of the position is far from the cloud of blue points, a corrected position is selected as the centroid of the votes weighted by image intensity. This process is repeated to correct errors for every neuron at every time.}
	\end{center}
\end{figure}

Neuron Registration Vector Encoding successfully identifies segmented neurons consistently across time. A transient segmentation error, however, would necessarily lead to missing or misidentified neurons. To identify and correct for missing and misidentified neurons, we check each neuron's locations and fill in missing neurons using a consensus comparison and interpolation in a thin-plate spline (TPS) deformed space. For each neuron identifier $s$ and time $t^\star$, we use all other point-sets, $\{\bfX_t\}$ to guess what that neuron's location might be. This is done by finding the TPS transformation, $u_{t\rightarrow t^\star}:\bfX_t \mapsto \bfX_{t^\star}$, that maps the identified points from $\bfX_t$ to the corresponding points in $\bfX_{t^\star}$ excluding the point $s$. The position estimate is then given by $u_{t\rightarrow t^\star}[\bfx_{i,t}]$ with $ i $ selected such that $ s_{i,t} = s$. This results in a set of points representing the predicted location of the neuron at time $t^\star$ from all other times. When a neuron identifier is missing for a given time, the position of that neuron $s$ is inferred by consensus.  Namely,   correct location is deemed to be the centroid of the set of inferred locations weighted by the underlying image intensity. This weighted centroid is also used if the current identified location of the neuron $s$ has a distance greater than 3 standard deviations away from the centroid of the set of points, implying that an error may have occurred in the assignment. This is shown in Fig \ref{correction}, where neuron 111 is identified in volume 735, but the the label for neuron 111 is incorrectly located in volume 736. In that case the weighted centroid from consensus voting was used.

\subsection{Comparison with manually annotated data}
\begin{figure}
	\begin{center}	
		 \includegraphics[width=.8\textwidth]{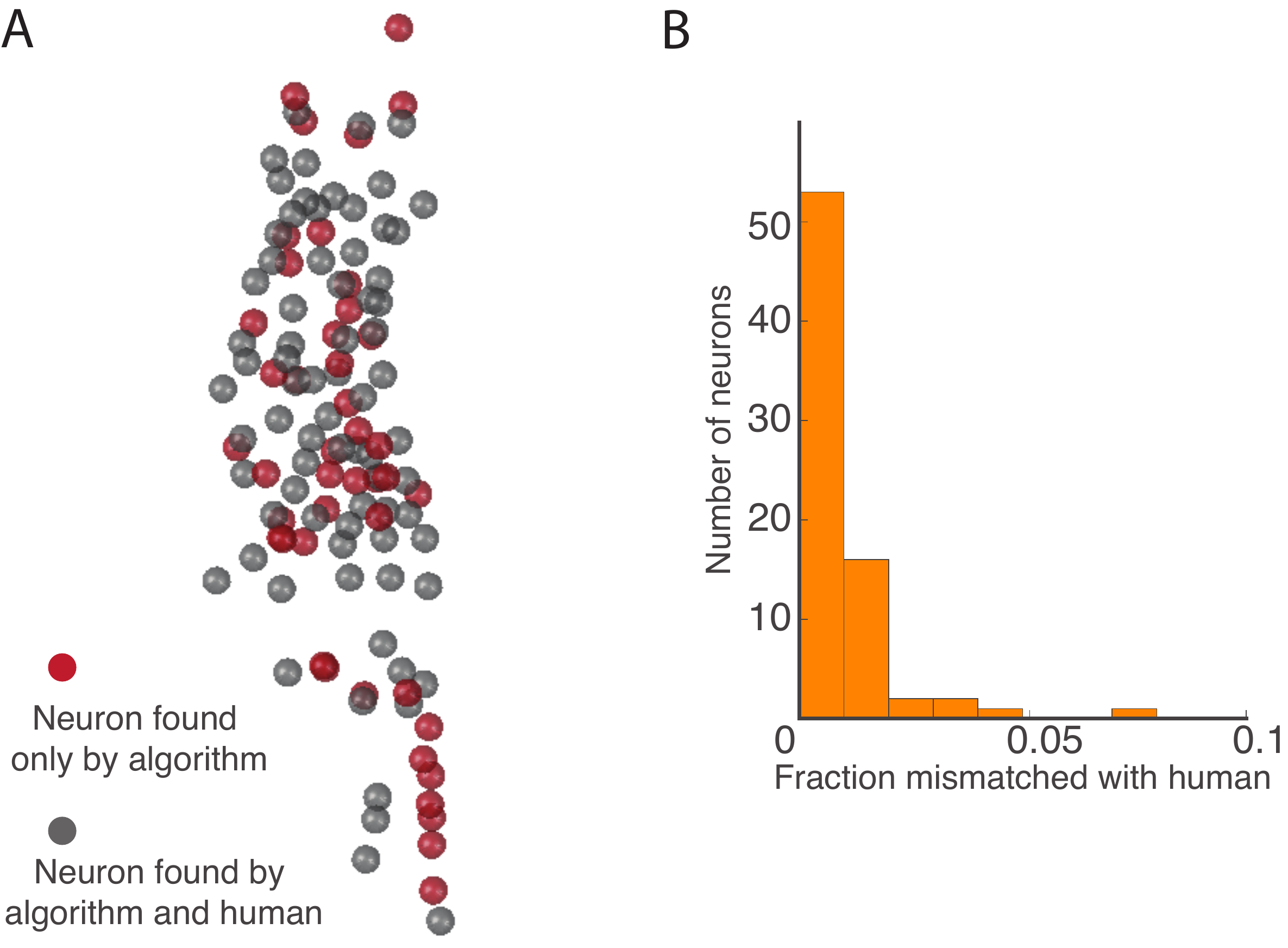}
	\caption{\label{errorRates} Comparison of the automated Neuron Registration Vector Encoding algorithm with manual human annotation over a 4 minute recording on brain activity (strain AML14).  (A)  Spheres show position of neurons  that were detected by the automated  algorithm.  Grey indicates a neuron detected by both the algorithm and the human (70 neurons). Red indicates neurons that were missed by the human and detected only by the algorithm (49 neurons). (B)  Histogram showing number of neurons that were mismatched for a given fraction of time-volumes when comparing automated and manual approaches.  Only those neurons that were consistently found by both algorithm and human were considered.  An automatically identified neuron  was deemed correctly matched for a given time-volume if it was paired with the correct corresponding manual neuron. }
	\end{center}
\end{figure}

To asses the accuracy of the Neuron Registration Vector Encoding pipeline, we applied our automated tracking system to a 4 minute recording of whole brain activity in a moving \textit{C. elegans}  that had previously been hand annotated \cite{Nguyen2016}. A custom Matlab GUI was used for manually identifying and tracking neurons. Nine researchers collectively annotated 70 neurons from each of the 1519 volumes in the 4 minute video. This is much  less than the 181 neurons predicted to be found in the head \cite{white1986structure}. The discrepancy is likely caused by a combination of imaging conditions and human nature. The short exposure time of our recordings  makes it hard to resolve dim neurons,  and the relatively long  recordings tend to cause photobleaching which  make the  neurons even dimmer. Additionally, human researchers naturally tend  to select only those neurons that  are brightest and are most unambiguous for annotation, and tend to skip dim neurons or those neurons that are most densely clustered. 

We compared human annotations to our  automated analysis in this same dataset. We  performed the entire pipeline including detecting centerlines, performing worm straightening, segmentation, and neuron registration vector encoding and clustering, and correction. Automated tracking detected 119 neurons from the video compared to 70 from the human.  In each volume, we paired the automatically tracked neurons with those found by manual detection by finding the closest matches in the unstraightened coordinate system. A neuron was perfectly tracked if it matched with the same manual neuron at all times. Tracking errors were labelled when a neuron matched with a manual neuron that was different than the one it matched with most often. The locations of the detected neurons are shown in Fig \ref{errorRates}A. Only one neuron was incorrectly identified in more than 5\% of the time volumes (Fig \ref{errorRates}B).   The locations of neurons and the corresponding error rates is shown in Fig \ref{errorRates}B. Neurons that were detected by the algorithm but not annotated manually are shown in gray.  Upon further inspection, it was noted that some of the mismatches between our method and the manual annotation were due to human errors in the manual annotation, meaning the algorithm is able to correct humans on some occasions.

\subsection{Neural activity from tracked neurons}
\begin{figure}
	\begin{center}	
		 \includegraphics[width=\textwidth]{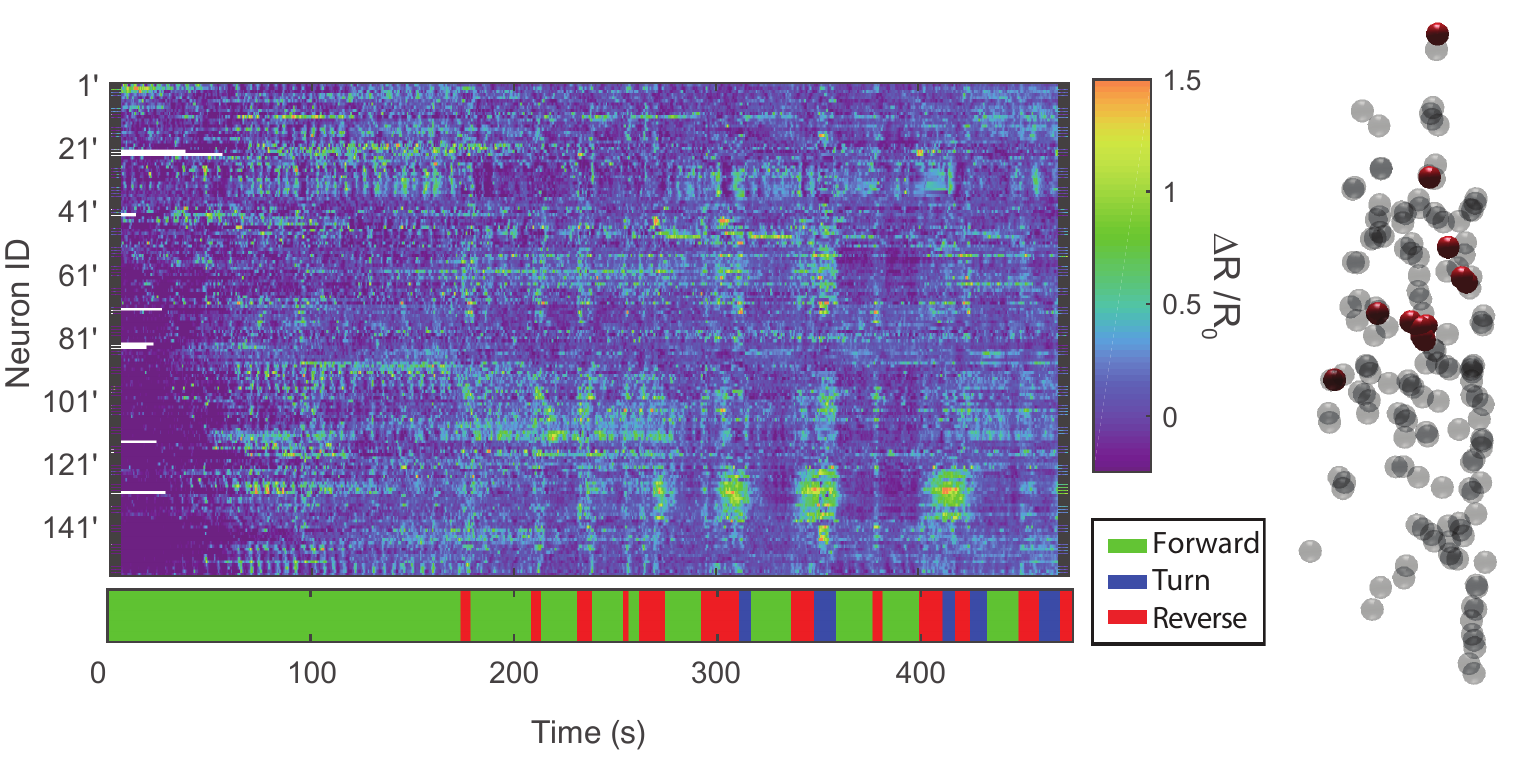}
	\caption{\label{heatmap} A trace of neural activity from 144 neurons in the brain  a \textit{C. elegans} as it freely moves  on an agarose plate over 8 minutes (strain AML32). The neural activity is expressed as a fold change over baseline of the ratio of GCaMP6s to RFP for each neuron. The behavior is indicated in the ethogram. On the right is the locations of all of the detected neurons (the head of the worm is towards the top of the page). The neurons that have significant correlation with reverse locomotion are indicated in red. White gaps indicate instances where neurons failed to segment. }
	\end{center}
\end{figure}

Fluorescent intensity is ultimately the measurement of interest and this can be easily extracted from the tracks of the neuron locations across  time. The pixels within a 2 $\mu m$ radius sphere around each point are used to calculate the average fluorescent intensity of a neuron in  both the  red RFP and green GCaMP6s channels. We used the RFP as a reference fluorophore and measure neural activity as a fold change over baseline of the ratio of GcAMP6s to RFP intensity,
\begin{align}
\mathrm{Activity}=\frac{\Delta R}{R_0} =\frac{R-R_0}{R_0},  &&\ R=\frac{I_{GCaMP6s}}{I_{RFP}}.
\end{align}
The baseline for each neuron, $R_0$, is defined as the 20th percentile value of the ratio $R$ for that neuron. Fig \ref{heatmap} shows calcium imaging traces extracted from new whole-brain recordings  using the registration vector pipeline. 144 neurons were  tracked for approximately 8 minutes as the worm moves. Many neurons show clear correlation with reversal behaviors in the worm.

\section{Discussion}

The Neuron Registration Vector Encoding method presented here is able to process longer recordings and locate more neurons with less human input compared to previous examples of whole-brain imaging in freely moving \textit{C. elegans} \cite{Nguyen2016}. Fully automated image processing means that we are no longer limited by the human labor required for manual annotation. In new recordings presented here, we are able to observe 144 neurons of the expected 181 neurons, much larger than the ~80 observed in previous work from our lab and others \cite{Nguyen2016,Venkatachalam2016}. By automating tracking and segmentation, this relieves one of the major bottlenecks to analyzing longer recordings.
%and in fact here we demonstrate a recording of  longer than 6 minutes, while previous methods demonstrated at most about 4 minutes of recording. %I don't know..l this just seemed kind of week.

The neuron registration vector encoding algorithm primarily relies on the local coherence of the motion of the neurons. It permits large deformations of the worm's centerline so long as  deformations around the centerline remain modest.  Crucially,  the algorithm's time-independent approach allows it to tolerate large motion between consecutive time-volumes. These properties make it well suited for our neural recordings of \textit{C. elegans} and we suspect that the our approach would be applicable to tracking neurons in moving and deforming brains from other organisms as well.  

Certain classes of recordings, however, would not be well suited for Neuron Registration Vector Encoding and Clustering. The approach will fail when the local coherence of neuron motion breaks down. For example, if one neuron were to completely swap locations with another neuron relative to its surroundings,  registration would not detect the switch and our method would fail. In this case a time-dependent tracking approach may perform better.

In addition, proper clustering of the feature vectors requires the animal's brain to explore a contiguous region of deformation space. For example, if a hypothetical brain were only ever to occupy two distinct conformations that are different enough that registration is not reliable between these two conformation states, the algorithm would fail to cluster feature vectors from the same neuron across the two states. To effectively identify the neurons in these two conformations, the animal's brain must sample many conformations in between those two states. This way, discrepancies in registration arise gradually and the resulting feature vectors occupy a continuous region in the space of possible feature vectors. Note that a similar requirement would necessarily apply to any time-dependent tracking algorithm as well.  

%Fortunately the  brains of most species,  including those from flies, rodents and primates should have the required properties of local coherence of neural motion 

 %mammals systems of interest 

% time  time-dependent tracking algorithm may h . Fortunatley for most organisms, neurons do not undergo diffusive motion.

 %  The algorithm ignores time and solely depends on the fact that the motion of the neurons is locally coherent and the deformations around the worm's centerline are not large. In systems where 3D imaging can be performed more rapidly, time information can be used to constrain the tracking and allow us to track the neurons in more situations.

%As it is, there are certain situations where fingerprinting and clustering would fail in the absence of additional information from time. If deformations are large or if neurons to completely switch positions, point-set registration will not be able to capture these deformations and the resulting fingerprints will not be able to identify certain neurons. 

%In addition, proper clustering of the fingerprints requires the animal to explore a continuous region in the space of deformations. For example, if the animal only ever occupies two distinct conformations that are different enough that registration is not reliable between these two states, clustering will not properly combine the fingerprints from the same neuron across the two states.  The animal must sample many conformations in between those two states so that any errors in the matches arise gradually and the resulting fingerprints can be clustered based on distances.

We suspect that  brain recordings from most species of interest meet these two requirements: namely neuron motion will have local coherence and the brain will explore a contiguous region of deformation space. Where these conditions are satisfied, we expect registration vector encoding to work well.  Tracking in \textit{C. elegans} is especially challenging because the entire brain undergoes large deformations as the animal bends. In most other organisms like zebrafish and \textit{Drosophila}, brains are contained within a skull or exoskeleton and relative motion of the neurons is small. In those organisms, fluctuations in neuron positions take the form of rigid global transformations as the animal moves, or local non-linear deformations due to motion of blood vessels. We expect that this approach will be applicable there as well.

\section{Methods}
\subsection{Strains}
Transgenic worms were cultivated on nematode growth medium (NGM) plates with OP50 bacteria. Strain AML32 (wtfIs5[P\textit{rab-3}::NLS::GCaMP6s; P\textit{rab-3}::NLS::tagRFP]) was generated by UV irradiating animals of strain AML14 (wtfEx4[P\textit{rab-3}::NLS::GCaMP6s; P\textit{rab-3}::NLS::tagRFP]) \cite{Nguyen2016} and outcrossing twice. 

\subsection{Imaging \textit{C. elegans}}
Imaging is performed as described in Nguyen et al \cite{Nguyen2016}. The worm is placed between an agarose slab and a large glass coverslip. The coverslip is held up by a 0.006'' plastic shims in order to reduce the amount of pressure on the worm from the glass, and mineral oil is spread over the worm to better match refractive indices in the space between the coverglass and the worm. The dark field image is used to extract the animal's centerline while the fluorescent image is used for tracking the worm's brain. Only the head of the worm is illuminated by the fluorescent excitation light and can be observed in the low magnification fluorescent image. 

\subsection{Thin Plate Spline deformations}
Point-set registration was done as described by Jian and Vemuri \cite{Jian2005}, using TPS deformations. Given a set of $n$ initial control points $\bfX=\{\bfx_i\}$, and the set of transformed points, $u[\bfX]$, the transformation $u$ can be written as $u[\bfX] = \mathbf{WU(X)+AX + t}$. The affine portion of the transformation is $\mathbf{AX + t}$, while $\mathbf{WU(X)}$ is the non-linear part of the transformation from TPS. $\mathbf{U(X)}$ is an $n \times 1$ vector with $\mathbf{U_i(x)=U(x,x_i)=U(\norm{x-x_i})=\frac{1}{\norm{x-x_i}}}$ and $\bfW$ is a $3 \times n$ matrix. The elements of $\bfW$, $\mathbf{A}$ and $\mathbf{t}$ can be fit given the set of control points $\bfX$ and the location of the transformed points $u(\bfX)$. The energy of bending, $E_{\mathrm{Bending}}(u)$, depends on how the control points are deformed. We use the same energy as in \cite{Jian2005}, with $E_{\mathrm{Bending}}(u)=\mathrm{trace}(\mathbf{WKW^T})$ where $\mathbf{K_{ij}}=U(\bfx_i,\bfx_j)$. Since the integral in Eqn.~\ref{Energy} is easily computed analytically, the energy can be quickly calculated and the parameters for $\bfW$, $\mathbf{A}$, and $t$ for the minimization can be found using gradient descent.

\subsection{Algorithm implementation}

The analysis steps shown in Fig \ref{flow} were performed on Princeton University's  high-performance scientific computing cluster, ``Della.'' Jobs were run on up to 200 cores. Straightening, segmentation, and feature extraction are parallelized over each volume, with each volume being processed on a single core. Error correction is parallelized over each neuron. Centerline extraction in each image relies on the previous centerline and must be computed linearly. The computation methods are summarized in Table \ref{table:compute}. An 8 minute recording of a moving animal has about 3000 volumes and 250 GB of raw imaging data and can be processes from start to finish in less than 40 hours. Data can be found via a ``requester-pays'' Amazon Web Services S3 ``bucket'' at   \url{ s3://leiferlabnguyentracking} and code can be found at \url{https://github.com/leiferlab/NeRVEclustering}. Centerline extraction code is available at \url{https://github.com/leiferlab/CenterlineTracking}. 
\begin{table}[]
\centering
\begin{tabular}{|l|l|l|}
\hline
Analysis Step                                                                & Computation & \begin{tabular}[c]{@{}l@{}}Approximate\\  Percentage of time\end{tabular} \\ \hline
Centerline Detection                                                         & Linear      & 4                                                                        \\ \hline
\begin{tabular}[c]{@{}l@{}}Worm Straightening\\ \& Segmentation\end{tabular} & Parallel over volumes    & 10                                                                       \\ \hline
\begin{tabular}[c]{@{}l@{}}Registration Vector\\ Encoding\end{tabular}       & Parallel over volumes   & 80                                                                       \\ \hline
Clustering                                                                   & Linear      & 2                                                                        \\ \hline
Error Correction                                                             & Parallel over neurons   & 4                                                                        \\ \hline
\end{tabular}
	\caption{\label{table:compute} Breakdown of computation time for Neuron Registration Vector Encoding pipeline.  }
\end{table}

\section{Acknowledgments}
% Acknowledgements for Plos Comp Bio:
%The analysis presented in this article was performed on computational resources supported by the Princeton Institute for Computational Science and Engineering (PICSciE) and the Office of Information Technology's High Performance Computing Center and Visualization Laboratory at Princeton University.
% For Plos Comp Bio Internal: This work was supported by Simons Foundation Grant SCGB 324285 to AML (https://www.simonsfoundation.org/life-sciences/simons-collaboration-global-brain/)  and Princeton University's Inaugural Dean for Research Innovation Fund for New Ideas in the Natural Sciences to JWS and AML (http://www.princeton.edu/main/news/archive/S39/38/59M52/index.xml?section=topstories).  JPN is supported by grants from the Swartz Foundation (http://www.theswartzfoundation.org/princeton.asp) and the Glenn Foundation for Medical Research (http://glennfoundation.org/glenn-centers/princeton/). ANL is supported by a National Institutes of Health institutional training grant NIH T32 MH065214 through the Princeton Neuroscience Institute. The funders had no role in study design, data collection and analysis, decision to publish, or preparation of the manuscript.

%Acknowledgements arXiv
We thank the following for their help manually annotating recordings: F Shipley, M Liu, S Setru, K Mizes, D Mazumder, J Chinchilla, and L Novak.   
This work was supported by Simons Foundation Grant SCGB 324285 (to AML) and Princeton University's Inaugural Dean for Research Innovation Fund for New Ideas in the Natural Sciences (to JWS and AML).  JPN is supported by a grant from the Swartz Foundation. ANL is supported by a National Institutes of Health institutional training grant through the Princeton Neuroscience Institute. This work is also supported by a grant from the Glenn Foundation for Medical Research. The analysis presented in this article was performed on computational resources supported by the Princeton Institute for Computational Science and Engineering (PICSciE) and the Office of Information Technology's High Performance Computing Center and Visualization Laboratory at Princeton University.

\section{Author contributions}
	%These catagories are specified and described here: http://journals.plos.org/ploscompbiol/s/authorship
   Conceptualization:  AML JWS JPN \\
   Methodology: JPN ANL \\
   Software: JPN ANL \\
   Validation: JPN \\
   Formal Analysis: JPN \\
   Investigation: JPN ANL \\
   Resources: JPN ANL GSP \\
   Writing - Original Draft Preparation: JPN\\
   Writing - Review \& Editing:  AML JWS\\
   Supervision: AML\\
   Funding Acquisition: AML\\

\bibliographystyle{unsrt}
\bibliography{PlosCombBio20150620}
\end{document}